\documentclass[traditabstract]{aa}
\usepackage{txfonts}
\usepackage[english]{babel}
\usepackage{ucs}
\usepackage[utf8x]{inputenc}
\usepackage{flushend}

\usepackage{graphicx}
\usepackage{wrapfig}
\usepackage{amstext}
\usepackage{amssymb}
\usepackage{setspace}

\usepackage{lscape}

\usepackage{natbib}
\bibpunct{(}{)}{;}{a}{}{,}

\usepackage{pgf}
\usepackage{pgfplots}
\usepackage{tikz}
\usepackage{url}
\usetikzlibrary{shapes,decorations.pathreplacing}

\usepackage{caption}
\usepackage{color}
\definecolor{lightgrey}{rgb}{0.7,0.7,0.7}
\definecolor{lightblue}{rgb}{0.5,0.5,1}
\definecolor{darkblue}{rgb}{0,0,0.5}
\definecolor{darkred}{rgb}{1,0,0}
\definecolor{olivegreen}{rgb}{0,0.5,0}
\definecolor{dunkelgrau}{rgb}{0.9,0.9,0.9}
\definecolor{hellgrau}{rgb}{0.95,0.95,0.95}
\usepackage{colortbl}

\begin{document}

\title{Prediction of Astrometric Microlensing Events during the\\ Gaia Mission}
\author{Svea Proft
\and Markus Demleitner
\and Joachim Wambsganss}
\institute{Astronomisches Rechen-Institut, Zentrum für Astronomie der Universität Heidelberg, Mönchhofstraße 12-14, 69120 Heidelberg, Germany} 
\date{Received 8 July 2011 / Accepted 30 September 2011}

\abstract {
We identify stars with large proper motions that are potential candidates for the astrometric microlensing effect during the Gaia mission i.e. between 2012 and 2019. The effect allows a precise measurement of the mass of a single star that is acting as a lens. We construct a candidate list by combining information from several input catalogs including PPMXL, LSPM, PPMX, OGLEBG, and UCAC3. The selection of the microlensing candidates includes the verification of their proper motions as well as the calculation of the centroid shift of the source resulting from the astrometric microlensing effect. The assembled microlensing catalog comprises 1118 candidates for the years 2012 to 2019. Our analysis demonstrates that 96\% of the (high) proper motions of these candidates are erroneous. We are thus left with 43 confirmed candidates for astrometric microlensing during the expected Gaia mission. For most of them the light centroid shift is below $\sim$100 µas (assuming a dark lens) or the astrometric deviation is considerably reduced by the brightness of the lens. Due to this the astrometric microlensing effect can potentially be measured for 9 candidates that have a centroid shift between 100 and 4000 µas. For 2 of these astrometric microlensing candidates we predict a strong centroid shift of about 1000 and 4000 µas, respectively, that should be observable over a period of a few months up to a few years with the Gaia mission.}
\keywords{astrometry - high proper motion stars - gravitational lensing}
\maketitle

\section{Introduction}
In the last three decades, gravitational lensing has become an important tool in astronomy and cosmology \citep{2006glsw.book.....S}. It is used e.g. to study the mass distribution of galaxies and clusters, to define the large scale geometry of the universe from the determination of the Hubble constant, to discover distant quasars \citep{1992ARA&A..30..311B} or to detect extrasolar planets \citep{1991ApJ...374L..37M}. The effect of gravitational lensing is the deflection and magnification of light from a background source by an intervening massive object. For the special case of stellar lenses and sources (Microlensing) there exist two images of the source. If observer, lens and source are perfectly aligned the image will be a ring with angular Einstein radius
	\begin{equation}
		\theta_\text{E}=\sqrt{\frac{4GM_\text{L}}{c^2}\frac{D_\text{L}-D_\text{S}}{D_\text{L} D_\text{S}}},
	\end{equation}
where $M_\text{L}$ is the mass of the lens and $D_\text{S},~D_\text{L}$ are the distances of the source and the lens to the observer \citep{1924AN....221..329C,1936Sci....84..506E,P86}. Typical values of the Einstein radius for Milky Way stars within one kpc distance as lens and distant sources are in the area of a few mas. If the components of gravitational lensing are closely, but not perfectly aligned the two images have an angular separation of order two Einstein radii. Hence it is not possible to resolve the source images with current telescopes. Because the images are not resolvable one can measure only their light centroid. Due to the relative motion of lens, source and observer, the magnification and image geometry changes with time hence the light centroid changes its position as well. The observation of the resulting light centroid shifts of source stars due to gravitational lensing is called astrometric microlensing. 

Bohdan Paczynski discussed the interesting possibility to determine the mass of a single star with astrometric microlensing \citep{astro-ph/9504099v1}. With this effect it would be possible to determine stellar masses with a precision of about one percent. Except for our sun and a few stars like MACHO-LMC-5 all known geometrically derived stellar masses are obtained from stars who are members of a double or multiple system \citep{2004ApJ...614..404G}. But double stars may not always evolve like single stars. Hence direct measurements of single star masses are very important.  

Another very nice feature of astrometric microlensing was emphasized by \citet{astro-ph/9504099v1}: Astrometric microlensing events can be predicted. For this, one has to identify faint stars with high proper motion. Faint stars are favored because the light centroid shift of the source star is less affected than with a bright lens. High proper motion stars are essential for a short enough time scale. Furthermore these fast moving stars are members of the solar neighborhood which allows the calculation of the lens distances from a parallax measurement with high accuracy. Additionally, the centroid shift is considerably larger for close lenses. Most known high proper motion stars are in a sphere with a radius of about 100 parsec. These stars of the solar vicinity are mainly faint Red Dwarfs. 

The first systematic search for nearby astrometric microlensing events was done by \cite{astro-ph/9909455v2} to identify candidates for the Space Interferometry Mission (SIM). They found 178 candidates during 2005 and 2015 with minimal lens-source angular distances between 7 and 20000 mas. Since then a couple of catalogs with high accuracy in position and proper motion containing faint stars were published, hence it is possible to find more astrometric microlensing candidates, and a calculation of the corresponding centroid shift is more reliable. 

With the Gaia mission it will be possible to measure the effect of astrometric microlensing with the required accuracy of order 10 to 100 µas. The satellite, expected to be launched in May 2013, surveys the whole sky and has a lifetime of five years \citep{ESAHP}. It will do astrometry, photometry and spectroscopy of approximately one billion stars in our Galaxy brighter than $\approx$20 mag in the visual band. The Gaia satellite will be the successor of Hipparcos \citep{1997yCat.1239....0E}, the first astrometric satellite. Between 1989 and 1993 Hipparcos measured about one million star positions with unrivaled precision of 1 to 20 mas for stars brighter than 12 mag.

An important quantity for the prediction of microlensing candidates is the astrometric accuracy $\sigma_\text{a}$ of Gaia. It will be $\sim$30 to 1400 µas for a single astrometric measurement for stars of brightness 10 to 20 mag \citep{astro-ph/0112243v1}. At the end of the Gaia mission the combined accuracy, which is not relevant for our study, varies between $\sim$5 and 200 µas depending on the stellar brightness and the actual number of measurements \citep{ESAPER}. The photometric precision for a single measurement will be $\sim$1 to 20 millimag \citep{2011EAS....45..167V,MIGNARD:2011:HAL-00602701:1,NewEntry4} and the photometric accuracy reached at the end of the mission will amount to $\sim$0.1 to 2 millimag for stars of brightness 10 to 20 mag \citep{NewEntry4,astro-ph/0101235v1,2011EAS....45..167V}, respectively.

In this paper we selected astrometric microlensing candidates during the Gaia mission. We explore the possibility that Gaia can measure the resulting centroid shift trajectories. 

The paper is structured as follows: In Section 2 we recall the photometric and astrometric signatures of microlensing and the mass determination of the lenses. In Section 3, we present our catalog ''Candidates for Astrometric Lensing'' and explain its features. In Section 4 we show the candidates for astrometric lensing that have the closest approach between 2012 and 2019 and discuss the candidates that should be observable with Gaia. Finally we summarize the results and give an outlook in Section 5.

\section{Basics of Microlensing}
	\subsection{Photometric Microlensing}
	The magnification due to the focusing of light from a pointlike source by a pointlike lens is called photometric microlensing. The magnification of both source images ($A_+,~A_-$) and the total magnification $A$ of the light centroid depends only on the impact parameter $u=(\theta_\text{L}-\theta_\text{S})/\theta_\text{E}$, where $\theta_\text{L}$ and $\theta_\text{S}$ are the angular positions of lens and source. The impact parameter is the projected distance between lens and source in units of $\theta_\text{E}$ . With this the total magnification is represented by \citet{P86} as
		\begin{equation}
			A(u)=|A_+|+|A_-|=\frac{u^2+2}{u\sqrt{u^2+4}}.
		\end{equation}

	The magnification for large impact parameter is approximated by \citep{DominikSahu2000}:
	\begin{equation}
		\mu(u)\simeq 1+\frac{2}{(u²+2)²}
	\end{equation}
	and the corresponding magnitude shift by
		\begin{equation}
			\label{equ::deltaMag}
			\Delta m=-\frac{5}{\ln{10}\cdot u^4},
		\end{equation}
%
	i.e. photometric microlensing is measurable only for small values of the impact parameter ($u\lesssim1$). Hence the timescale of a photometric microlensing event is typically the Einstein time, the time to cross the Einstein radius, which is expressed by
		\begin{equation}
			\label{equ::einsteintime}
			t_\text{E}=\frac{D_\text{L}\theta_\text{E}}{v_\perp}= \frac{\theta_\text{E}}{\mu},
		\end{equation}
	where $\mu$ is the proper motion and $v_\perp$ the transverse velocity of the lens. From photometric measurements it is possible to derive $t_\text{E}$, but $D_\text{L}$, $v_\perp$ and $\theta_\text{E}$ (lens mass) can not be extracted individually from observations without additional information. 
	\subsection{Astrometric Microlensing}
	The astrometric microlensing effect of a close lens with $D_\text{L}=$50 pc and a distant source is illustrated in the upper panel of Figure 1 (following \citealt{astro-ph/9604011v1}). The lens has an Einstein radius of 8.5 mas and we assume the minimal impact parameter $u_0$ to be 0.5, which corresponds to a projected distance of 4.25 mas. When a high proper motion star (red crosses) is moving in front of a background source star (big black point) there are two image trajectories due to gravitational lensing (blue open circles). Even if the images are not resolvable one may measure a centroid shift trajectory (violet points) which is always an ellipse. Lens, source, images and the centroid position are all lying on a straight line (dashed line in Figure 1). The black circle presents the corresponding Einstein ring of the event connected with the dashed line. The lower image of Figure 1 shows different centroid trajectories of this astrometric microlensing event. The ellipses correspond to minimal impact parameter of $u_0=0.5$ (violet), 1 (blue), $\sqrt{2}$ (red), 2 (green), and 5 (black).
	\begin{figure}[tbp]
		\includegraphics[width=0.49\textwidth]{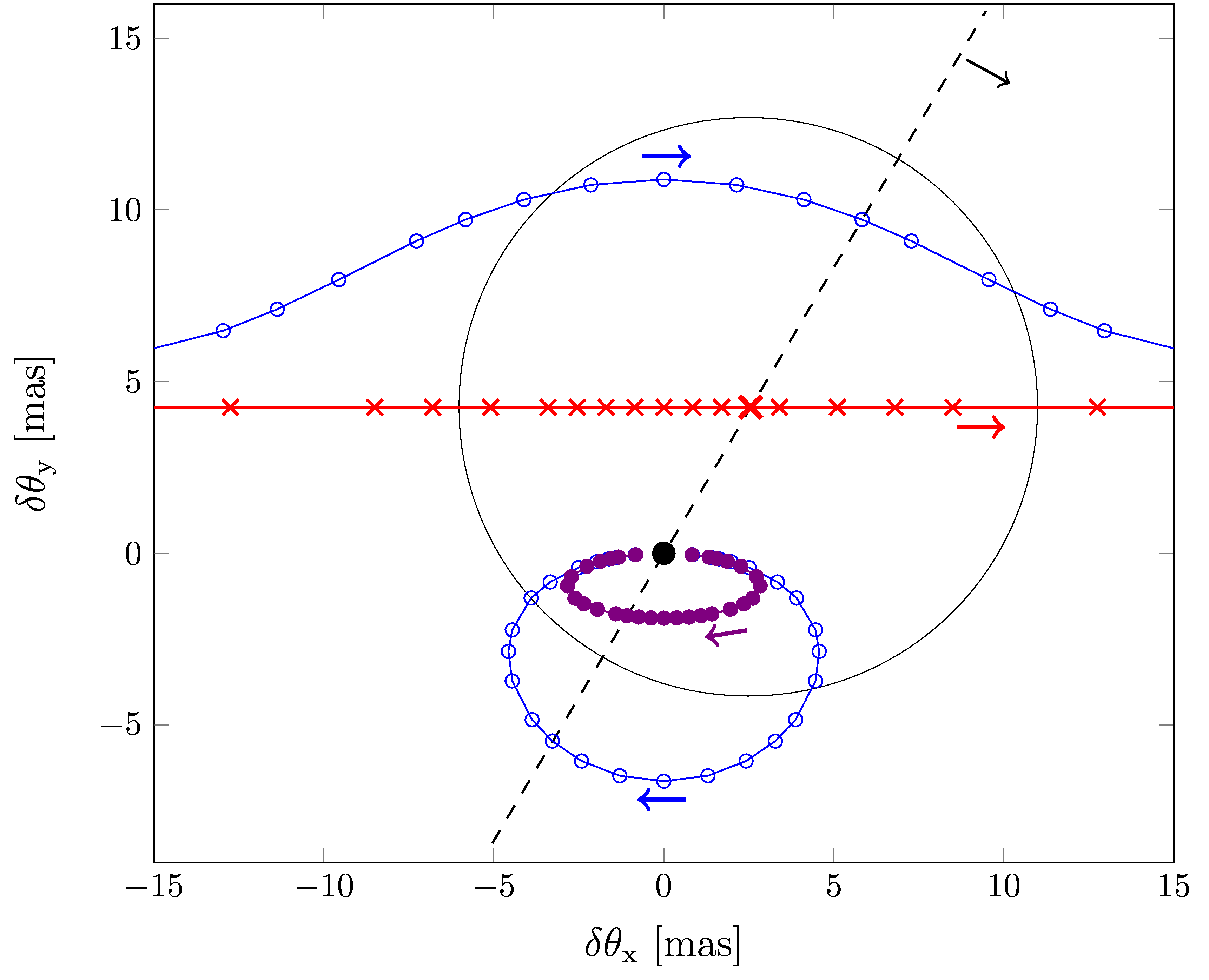}
		\includegraphics[width=0.48\textwidth]{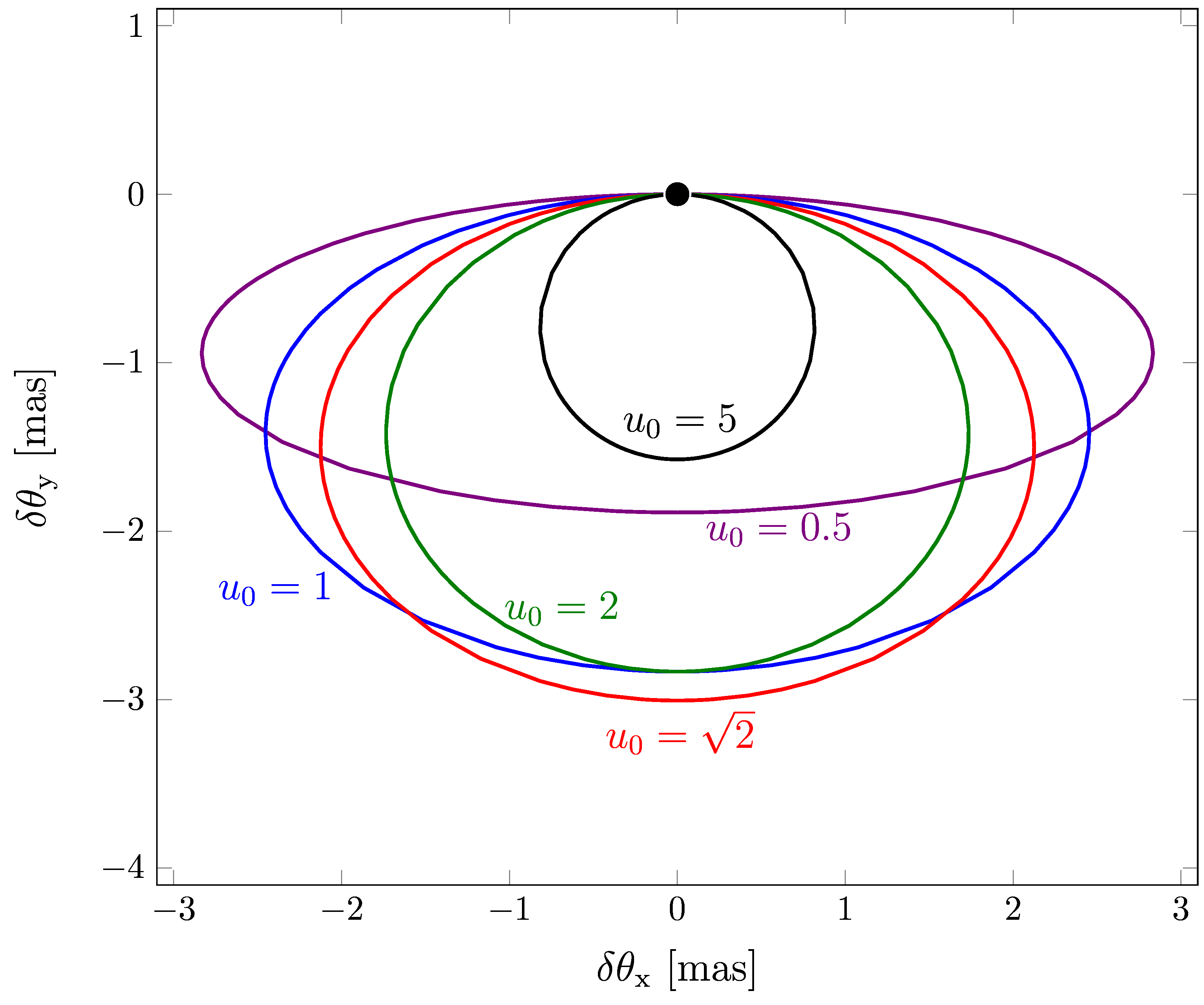}
    \captionsetup{format=plain}
		\vspace{-0.3cm}
		\caption{Upper panel: An astrometric microlensing event illustrated with a close lens ($D_\text{L}=50$pc, $\theta_\text{E}=8.5$mas) and a distant source (according to Figure 3 of \citealt{astro-ph/9604011v1}). The lens (red crosses) is moving in front of a background star (big black point) with $u_0=0.5$ (equivalent to 4.25 mas). Due to gravitational lensing there are two images whose paths are shown with blue open circles. If the two images are not resolvable one can measure only the light centroid shift trajectory (violet points) which is an ellipse. Lower panel: Different centroid shift trajectories of the same astrometric microlensing event but for different impact parameters $u_0=0.5$, 1, $\sqrt{2}$, 2 and 5 (according to Figure 2 of \citealt{DominikSahu2000}).}
    \end{figure}

	Here we explain the astrometric microlensing signals. First we assume a dark lens. The centroid position is given by \cite{1995A&A...294..287H}, \cite{1995AJ....110.1427M}, \cite{1995ApJ...453...37W}, and \cite{1005.3021v2} as
		\begin{equation}
			\vec\theta_\text{c}=\frac{A_+\vec\theta_+ + A_-\vec\theta_-}{A_+ + A_-},
		\end{equation}
	where $\theta_+$ is the image position outside and $\theta_-$ inside of the Einstein ring radius. The light centroid shift in comparison to the unperturbed source position
		\begin{equation}
			\label{equ::DeltaThetaC}
			\delta\vec\theta_\text{c}=\frac{\vec u}{u^2+2}\cdot\theta_\text{E}
		\end{equation}
	depends on the impact parameter and the Einstein radius. The inner dependence on $\theta_\text{E}$ is the reason why astrometric microlensing leads to a preference of nearby stars as lenses and distant sources. The corresponding maximal value of centroid shifts is determined by \citet{astro-ph/9708155v2} to
		\begin{equation}
			\delta\theta_\text{c,max}\approx0.354~\theta_\text{E}
		\end{equation}
	at $u=\sqrt2$. For instance, the maximal possible centroid shift of a one solar mass lens at 50 pc and a source at 5 kpc distance amounts to 4.5 mas.

	For large impact parameter ($u\gg\sqrt{2}$) one gets a centroid shift of \citep{DominikSahu2000}
		\begin{equation}
			\delta\theta_\text{c}\simeq \frac{\theta_\text{E}}{u},
		\end{equation}
	which implies that the centroid shift decreases only with $u^{-1}$ in comparison to the magnitude shift which depends on $u^{-4}$. Hence the cross section for astrometric microlensing is much larger than for photometric lensing \citep{astro-ph/9606060v1,astro-ph/9605138v1}.

	With consideration of the luminous lens effect the centroid position, 
	\begin{equation}
		\vec\theta_\text{c,lum}=\frac{A_+\vec\theta_+ + A_-\vec\theta_- + f_\text{LS}\vec\theta_\text{L}}{A_+ +A_-+f_\text{LS}},
	\end{equation}
	depends additionally on the flux ratio $f_\text{LS}=f_\text{L}/f_\text{S}$ of the lens and the source star. For u$\gg\sqrt{2}$ the centroid shift is reduced by a factor of $1+f_\text{LS}$ in comparison to a dark lens \citep{DominikSahu2000}: 
		\begin{equation}
			\label{equ::CenShiftMaxLum}
			\delta\theta_\text{c,lum}=\frac{1}{(1+f_\text{LS})u}\theta_\text{E}.
		\end{equation}
	If lens and source are both resolved, the luminous lens effect does not need to be considered anymore. The components are resolvable when their projected distance is larger than Gaia's angular resolution, which is about 200 mas.\\ 

	It is assumed that approximately half of all stars are binaries or multiple systems. In this case the centroid shift ellipse is characterized by distortions, twistings, and jumps \citep{1999ApJ...526..405H}. The majority of double stars in the solar neighborhood will be widely separated binary systems. This means they will have a projected binary separation $d$ much larger than the Einstein radius of the primary lens\footnote{The primary lens is defined as the binary component with the smallest angular separation to the source.}. The resulting light curve of this binary system will be very similar to the light curve of the primary lens. But the second lens component can affect the centroid shift trajectory to much larger projected distances (up to $\sim$100 $u_0$ depending on the mass of the second lens). For large projected binary separations and a source close to the primary lens the magnification and the centroid shift of the second lens component is approximated to \citep{2002ApJ...573..351A}
	\begin{equation}
		A_2\sim1+2\frac{q^2}{d^4} ~~~ (d\gg2),
	\end{equation}
	\begin{equation}
		\label{equ::deltaTheta2}
		\delta\theta_2\sim\theta_\text{E,1}\frac{q}{d} ~~~ (d\gg\sqrt{2}), 
	\end{equation}
	where $q=m_2/m_1$ is the mass ratio of the lens components. The centroid shift of a wide binary lens system can be expressed as the superposition of the individual centroid shifts of both components \citep{2009ApJ...695.1357C}:
	\begin{equation}
		\delta\vec\theta_\text{c}\sim\delta\vec\theta_\text{c,1}+\delta\vec\theta_\text{c,2}.
	\end{equation}

	Due to a larger cross section for astrometric lensing, the duration of astrometric microlensing events is much longer than for the corresponding photometric event. The average duration in which Gaia could measure the astrometric deviation is given by \cite{astro-ph/0102242v1} as
	\begin{equation}
		\label{equ::t_ast}
		t_\text{ast}=\frac{\pi}{2}\left(\frac{t_\text{E}\theta_\text{E}}{\theta_\text{min}}\right),
	\end{equation}

	\noindent\normalsize
	where $\theta_\text{min}$ is the accuracy threshold of Gaia for which the light centroid shift is measurable. We can estimate this threshold to be
	\begin{equation}
		\theta_\text{min}=\frac{3\sqrt{2}\sigma_\text{a}}{\sqrt{5}}
	\end{equation}
	where $\sigma_\text{a}$ is the astrometric accuracy of a single measurement, which depends on the source star brightness. The estimation for $\theta_\text{min}$ is a result of assuming the mean number of consecutive observations to be five (Bastian, priv. comm.), a 3-sigma-area to have a convincing centroid shift measurement, and to get a two-dimensional accuracy (in general Gaia has a high precision in only in one direction) \citep{astro-ph/0112243v1,2011EAS....45..109L}. 

	The mass of the lensing star can be determined \citep{astro-ph/9708155v2} as
		\begin{equation}
			\label{equ::lensmass}
			M_\text{L}=0.123~M_\odot \frac{\theta^2_\text{E}}{\pi_\text{LS}},
		\end{equation}
	where $\pi_\text{LS}=\pi_\text{L}-\pi_\text{S}$ is the parallax of the lens source system. By the use of astrometry it is possible to specify the parallax of at least the (visible) lensing star. When the lensing star is observable, the angular distance of lens and source is determinable. With additional information of the unaffected source position the centroid shift is measured. With given angular lens-source distance and centroid shift the Einstein radius is calculated by Equation (\ref{equ::DeltaThetaC}). Hence the mass of the lens could be computed by astrometry but the highest accuracy is achieved when the microlensing event is observed both astrometrically and photometrically. Therefore, an accompanying high-cadence ground based photometric monitoring of the events would be advantageous. If the lens is not observable, the combination of photometric and astrometric measurements is necessary. When Gaia will measure two sequenced positional shifts of a source star there will be an alert to do photometric observations from Earth (Wyrzykowski, priv. comm.; \citealt{NewEntry2}). For high proper motion stars it is likely that the time at closest approach, where a photometric event could be observable, has passed by that time. Hence it is important to predict microlensing events in particular for high proper motion lenses. 
\section{The Microlensing Catalog}
	\subsection{Prediction of Microlensing Events} 
	For the determination of microlensing candidates we search for (background) stars that lie within a certain angular distance to the future position of the high proper motion star as illustrated in Figure 2. The image demonstrates the path of a lensing star between 2008 and 2020. On both sides of the trajectory, a band is defined with a certain minimal angular width $w$ and four corner points. If a source star lies inside this area the lens will be assigned as a candidate for microlensing. In consideration of the positional and proper motion accuracies of the lenses and sources, $w$ was selected to be 0.7''. Typical accuracies are $\sim$100 mas in position and $\sim$10 mas/year in proper motion on the standard epoch 2000.0. The increasing width of the trapezium (dashed lines) reflects the increasing uncertainties in the lens position and proper motions as one moves away from this catalog epoch. However, it is sufficient to use the rectangle. Our list ''Candidates for Astrometric Lensing'' contains the results of this algorithm and can be found at the GAVO data center\footnote{http://dc.g-vo.org/aml}. It comprises about 2400 microlensing candidates that have their closest approach between 1950 and 2100.

	For the prediction of microlensing candidates it was essential to have a reliable catalog of proper motions. The currently most suitable catalog for the sources is the PPMXL \citep{1003.5852v1}. It is an all sky catalog with astrometric and photometric information of nearly one billion objects and has a limiting magnitude of $\approx$20 mag in the visual band. 

	\begin{figure}[t]
			\centering
			\includegraphics[width=0.47\textwidth]{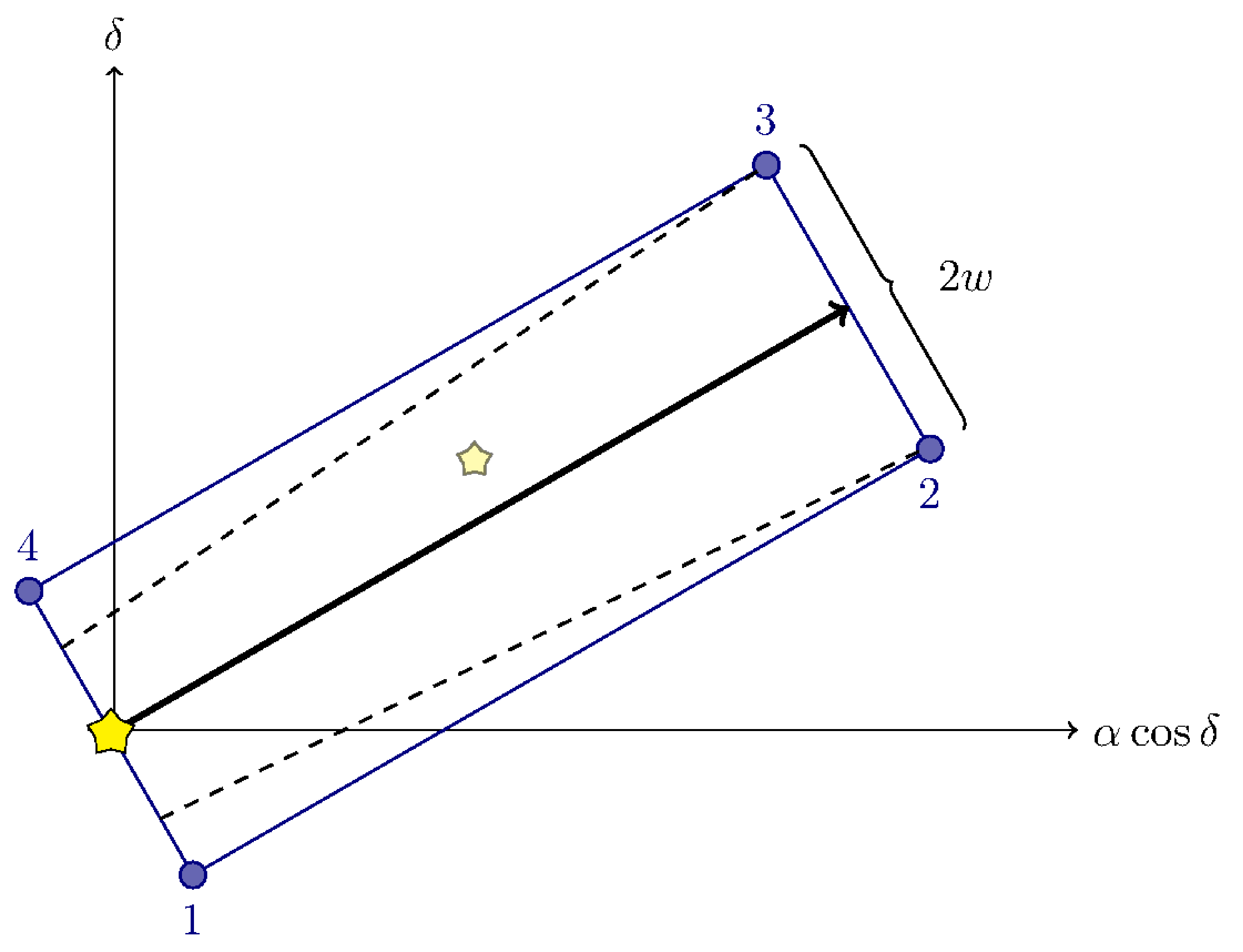}
			\captionsetup{format=plain}
			\vspace{-0.1cm}
			\caption{\label{abb::Candidates}Trajectory of a lensing star in equatorial coordinates between 2008 and 2020 (black arrow). If a background source star (small star) lies inside the area defined with a minimal angular width $2w$ and the corner points 1 to 4, the lens (big star) will be identified as candidate in our table ''Candidates for Astrometric Lensing''. The rectangle is a simplification of the trapezium (dashed lines), which expresses the increasing uncertainty in the lens positions and proper motions from the catalog epoch 2000.0.}
	\end{figure}

	In common astronomical catalogs the majority of stars with high proper motions ($\gtrsim0.15''$/year) are spurious. There were many problems with digitalization of old photographic plates, grains on them which were identified as stars, plates from different epochs in different filters or incorrect matching of star positions from different sources especially in regions with high star density. Hence it is difficult to find reliable catalogs for the lensing stars.

	Gaia's predecessor Hipparcos produced the Hipparcos and Tycho catalogs which consist of stars brighter than 12 mag. Stars in the solar neighborhood are in general much fainter than this limit. Hence Hipparcos and Tycho contain only few high proper motion stars. Furthermore there exist also some older high proper motion star catalogs like NLTT \citep{1980PMMin..55....1L} or LHS \citep{1979lccs.book.....L}, but their positions and proper motions are not precise enough for our study. Considering the limiting magnitude and the positional and proper motion accuracies we selected the LSPM-NORTH \citep{astro-ph/0412070v1}, PPMX \citep{0806.1009v1}, UCAC3 \citep{1003.2136v1}, and the "catalogue of stellar proper motions in the OGLE-II Galactic bulge fields" (OGLEBG) from \cite{2004MNRAS.348.1439S} as lens catalogs. Since the LSPM-NORTH is not available for the whole sky and the PPMX and UCAC3 have not the same quality in every part of the sky they were chosen for different sky regions. The limiting V magnitudes are $\approx$21mag (LSPM-NORTH), $\approx$15mag (PPMX) and $\approx$16mag (UCAC3). Because the probability for microlensing events increases with the density of source stars the majority of microlensing candidates should be close to the Galactic plane. Due to crowding the PPMX, UCAC3, and common proper motion catalogs have a lot of spurious proper motions in sky regions with high star density. Hence we also selected OGLEBG as a lens catalog. OGLEBG contains 5,080,236 stars towards the Galactic bulge with Galactic longitudes $-11°<l<11°$ and latitudes $-6°<b<3°$. Of those, about 100 stars have proper motions larger than 0.15''/yr.

	In the next section we will show that LSPM-NORTH is a suitable lens catalog. It is available only for the northern sky. This catalog contains nearly 62,000 stars with proper motions larger than 0.15''/year. The image technique of LSPM-NORTH is different from common proper motion catalogs and leads to correct star identifications and hence to correct proper motions. Another suitable lens catalog is the rNLTT of \cite{2003ApJ...582.1011S}. The rNLTT contains about 36,000 stars with declinations larger than -30°. The rNLTT proper motions are with $\sim$ 5.5 mas/year more accurate than the corresponding LSPM-NORTH proper motions ($\sim$ 8 mas/year). The rNLTT is the most precise high proper motion catalog with reliable proper motions and comprising faint stars, but it is less complete than LSPM-NORTH in the northern sky, hence we selected the LSPM-NORTH. We will consider the southern part of the rNLTT as lens catalog in future studies. 

	Most of our lensing stars are also contained in PPMXL, but occasionally their high proper motion is not listed correctly in this catalog. If the proper motion is known correctly in PPMXL and the proper motion accuracy is better than the considered accuracy of the lens catalog, we used the improved astrometry from PPMXL for the calculation of the centroid shift and the related quantities. The proper motion accuracy of PPMXL varies between 4 and 10 mas/year. The PPMXL mean errors of positions at standard epoch 2000.0 are 80 to 120 mas, when 2MASS astrometry was available \citep{1003.5852v1}.  

	With LSPM-NORTH, PPMX, OGLEBG and UCAC3 as lens catalogs and the PPMXL as source catalogs we find 1118 supposed astrometric microlensing candidates that have their closest approach between 2012 and 2019. Because many of the high proper motions are erroneous we have to individually check the proper motions of all supposed lensing candidates. 
	\subsection{High Proper Motion Analysis} 
		We began the high proper motion analysis with the method of reduced proper motions. This is an important tool to classify star populations \citep{astro-ph/9909455v2,astro-ph/0412070v1}. By this method a visual magnitude H$_\text{V}$ is estimated by use of the proper motion $\mu$. The absolute magnitude is given by 
		\begin{equation}
			M_\text{V}=m_\text{V}+5\log\pi+5.
		\end{equation}
		When the parallax $\pi$ is replaced by $\mu$, the reduced proper motion magnitude is \citep{1972ApJ...177..245J}
		\begin{equation}
			H_\text{V}=m_\text{V}+5\log\mu+5.
		\end{equation}

\begin{figure}[tbp]
		 	\centering
			\includegraphics[width=0.5\textwidth]{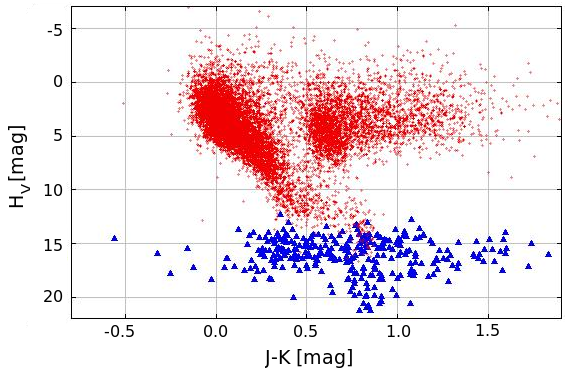}
			\captionsetup{format=plain}
			\vspace{-0.5cm}
			\caption{HRD produced with the method of reduced proper motions of the supposed microlensing candidates (blue triangles) and 10000 stars of the Hipparcos catalog (red dots). The magnitude $H_\text{V}$ is plotted against the color J-K. According to this plot all lenses should be a M-, Sub- or White Dwarf, but the distribution is very diffuse.}
			\label{fig::Luyten3}
		\end{figure}

		Most infrared magnitudes J,H and K are listed in our lens catalogs or are available from 2MASS \citep{2003tmc..book.....C}. Hence we can plot a reduced proper motion diagram, i.e. an Hertzsprung-Russell diagram (HRD) with H$_\text{V}$ against e.g. J-K. Figure \ref{fig::Luyten3} shows this diagram for our 1118 supposed microlensing candidates with known infrared colors in blue triangles and and a selection of 10000 Hipparcos stars in red dots. The Hipparcos stars (V$\lesssim$12) indicate the main sequence and giant branch. Our candidates are all in the area of M-Dwarfs, Subdwarfs or White Dwarfs, but the distribution is much more diffuse than for the Hipparcos stars.

		\begin{figure}[tbp]	\vspace{0.5cm}
			\centering
			\includegraphics[width=0.5\textwidth]{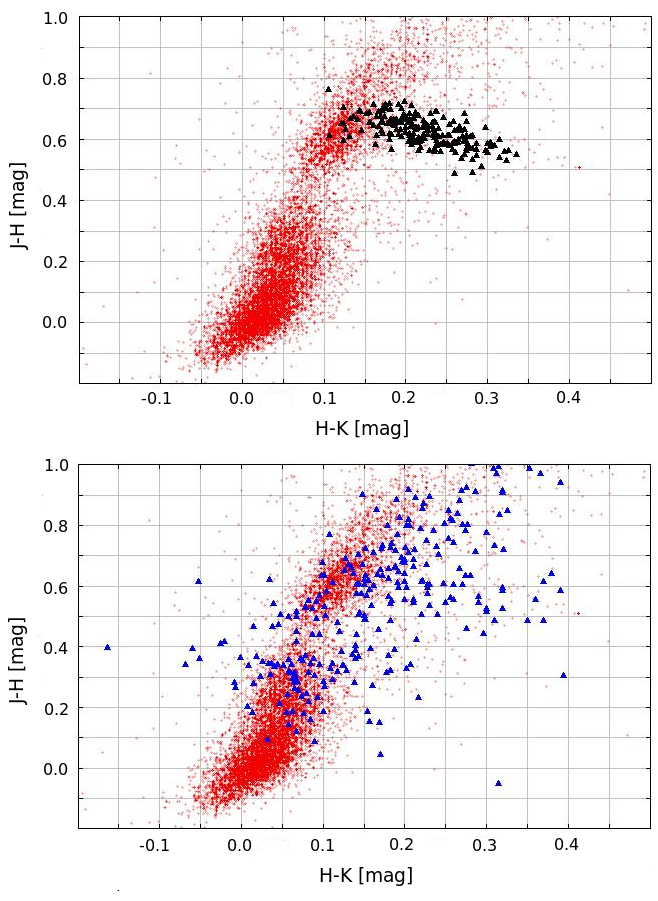}
			\vspace{-0.5cm}
			\captionsetup{format=plain}
			\caption{Two-color-diagram to characterize the microlensing candidates. The colors J-H against H-K are shown. In the upper panel 10,000 Hipparcos stars (red dots) and about 200 dwarfs (black triangles) from the "All-Sky Catalog of Bright M Dwarfs" \citep{1538-3881-142-4-138} are plotted. The lower panel presents the supposed microlensing candidates (blue triangles) and the same Hipparcos stars as in the upper panel.}
		\end{figure}

		The second proper motion analysis was done with a two color diagram (colors J-H against H-K). The upper panel of Figure 4 shows a two color diagram for 10,000 Hipparcos stars (red dots) and about 200 stars (black triangles) of the "All-Sky Catalog of Bright M Dwarfs" \citep{1538-3881-142-4-138}. The absolute magnitude of dwarfs and giants increases (fainter stars) from the bottom left to the upper right and the region at J-H$\approx$0.6 mag is a branch of M-Dwarfs \citep{BB88}. According to Figure \ref{fig::Luyten3}, the majority of our 1118 candidates should be in this M-Dwarf branch. The lower panel of Figure 4 shows the same two color diagram but for the selection of Hipparcos stars (red dots) and our microlensing candidates (blue triangles). One can see that the candidates are distributed over all stellar populations. A comparison of the results of the reduced proper motion diagram (Figure \ref{fig::Luyten3}) and the two color diagram (Figure 4) yields a discrepancy. Because the reduced proper motion diagram (Figure \ref{fig::Luyten3}) was constructed with the lens proper motions, this discrepancy shows that indeed the majority of lens proper motions are erroneous. Hence we have to check the 1118 microlensing candidates individually to estimate the correct proper motion values. 

		The analysis of the lens proper motions was done with the software sky atlas Aladin \citep{2000AAS..143...33B}. This means we had to ''blink'' like Lépine \& Shara for the generation of the LSPM-NORTH. In order to do that, we chose images from different sky surveys and estimated the lens proper motion. Investigating about 100 lens proper motions, we found that typically the proper motions were correct when the lenses were included in a high proper motion star catalog (like NLTT or LHS). For the remaining supposed candidates it was sufficient to check with SIMBAD \citep{2000AAS..143....9W} which of them are characterized as high proper motion stars.

		We found out that 96\% of the analyzed 1118 high proper motions are incorrect. Indeed, all LSPM-NORTH proper motions are correct, but most high proper motions from PPMX and UCAC3 were erroneous. We refer to \citet{1003.5852v1} for a discussion of a hugh number of high proper motion stars in the PPMXL. They pointed out that the majority of them must be wrong based on a comparison to a flat proper motion function between 150 and 430 mas/year. 

\section{Astrometric Microlensing Candidates}
Our list of high proper motion stars as candidates for astrometric microlensing with closest approach to a background source star between 2012 and 2019 comprises 1118 supposed candidates. After double checking the proper motions only 43 actual microlensing candidates remain. Out of those, 36 lensing candidates are on the northern and 7 on the southern sky. The lenses have proper motions between 0.17 and 2.38''/year and apparent magnitudes between 12.1 and 19.9 mag.

When we analyze the reduced proper motion diagram for the actual microlensing candidates (Figure 5) we can see again that our candidate lens stars that have known infrared colors (39 of 43) are all in the regions of Red or White Dwarfs. In comparison to Figure \ref{fig::Luyten3}, the distribution is very concentrated in a small area. In Figure 6, the two color diagram for the actual microlensing candidates with known infrared colors is shown. In this diagram the remaining candidates (blue triangles) are all located in the Red or White Dwarf branch. Now the comparison of the reduced proper motion diagram (Figure 5) and the two color diagram (Figure 6) of the 39 of 43 microlensing candidates with known infrared colors are in accord.

	\begin{figure}[h]	
		\includegraphics[width=0.5\textwidth]{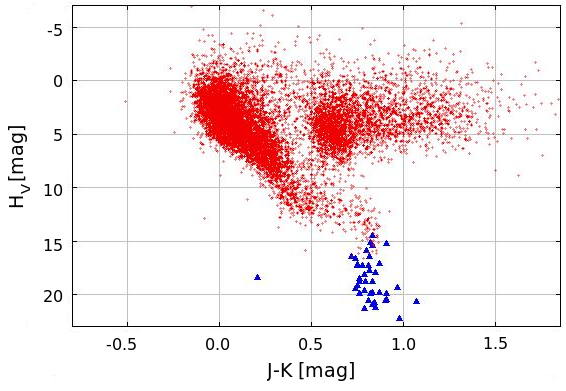}
		\centering
		\captionsetup{format=plain}
		\vspace{-0.5cm}
		\caption{Reduced proper motion diagram of the Hipparcos stars (red dots) and the 39 of our remaining 43 microlensing candidates with known infrared colors (blue triangles). All actual candidates are Red or White Dwarfs}
	\end{figure}

	\begin{figure}[tbt]
		\includegraphics[width=0.5\textwidth]{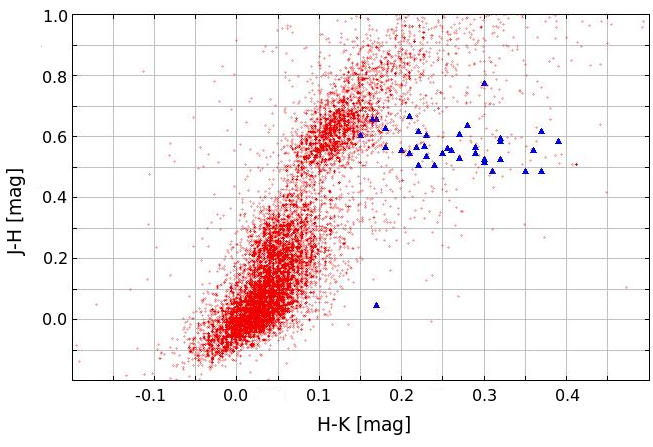}
		\centering
		\captionsetup{format=plain}
		\vspace{-0.5cm}
		\caption{Two color diagram of the Hipparcos stars (red dots) and our 39 real microlensing candidates with known infrared colors (blue triangles). All candidates are in the Red- or White Dwarf branch.}
	\end{figure}

Our 43 real microlensing candidates are listed in Table 1, ordered by time of closest approach. The table shows the lens ID, the lens and source right ascension and declination at epoch 2000.0, the apparent magnitudes of lens and source, the lens proper motion, the time of closest approach, and the minimal projected distance of lens and source. When we compare these 43 candidates with the 178 candidates from \cite{astro-ph/9909455v2}, who consider a different time interval, we can identify 4 lenses that are contained in both lists. These candidates are our candidate \#18 with Salim \& Gould \#45, \#19 with S. \& G. \#50, \#23 with S. \& G. \#137, and \#24 with S. \& G. \#18. For the 2nd of these candidates the event date is nearly the same (only 20 days difference) and the minimal angular distance only differs by $\approx$180 mas. For the other candidates the event date differs by more than 3 years.\vspace{0.2cm}

In order to find out which of our 43 candidates are actually detectable with Gaia, a few additional criteria have to be fulfilled:
\begin{itemize}
\item The maximal centroid shift has to be larger than the accuracy threshold $\theta_\text{min}$ of Gaia which depends on the source star brightness.
\item The astrometric event duration should be at least a few weeks depending on the time interval of observations with Gaia. On average, every two months there are a few consecutive measurements of every object within a few hours.
\item The source should be a star and not an extended object like a galaxy.
\end{itemize}

For the calculation of the maximal centroid shift we need to know the minimal projected distance of the involved stars and the angular Einstein radius. We have computed this angular distance and the time at closest approach with the corresponding uncertainties. To determine the Einstein radius we require an estimation of lens mass, lens distance and source distance. 4 of 43 lenses have known trigonometric parallaxes. For the others we had to resort to photometric parallaxes. Therefore we employ given infrared colors and the luminosity class (see Figures 5 and 6) to get the correct spectral type. Thus we could calculate the absolute magnitudes of our candidates. Then the distance is determined by the distance modulus. The calculated lens distances are between 6 and 476 pc. The lens mass is estimated using the

\newpage\onecolumn
\begin{table}\setstretch{1.1}
\begin{tabular}{|l|l|c|c|c|c|c|c|c|c|c|}
\hline
  \multicolumn{1}{|c|}{\#}&
  \multicolumn{1}{c|}{lens ID}&
  \multicolumn{1}{c|}{$\alpha_\text{L}$}&
  \multicolumn{1}{c|}{$\delta_\text{L}$}&
  \multicolumn{1}{c|}{$\mu_\text{L}$}&
  \multicolumn{1}{c|}{$m_\text{V,L}$}&
  \multicolumn{1}{c|}{$\alpha_\text{S}$}&
  \multicolumn{1}{c|}{$\delta_\text{S}$}&
  \multicolumn{1}{c|}{$m_\text{V,S}$}&
  \multicolumn{1}{c|}{$t_\text{min}$}&
  \multicolumn{1}{c|}{$d_\text{min}$}\\
&&(deg)&(deg)&(''/yr)&(mag)&(deg)&(deg)&(mag)&(yr)&('')\\
\hline\hline
\rowcolor{dunkelgrau}
\textbf{1}&\textbf{UCAC3  206.42273}	&\textbf{206.42273}	&\textbf{-51.01684}	&\textbf{0.527}	&\textbf{15.3}	&\textbf{206.42335}	&\textbf{-51.01857}	&\textbf{18.5}	&\textbf{2012.11}	&\textbf{0.070}\\
&\cellcolor{dunkelgrau}~~~~~~\textbf{-51.01684}	&	&	&	&	&	&	&	&	&\\
\rowcolor{dunkelgrau}
\textbf{2}&\textbf{LSPM J2226+2913}	&\textbf{336.73694}	&\textbf{29.22514}	&\textbf{0.230}	&\textbf{13.6}	&\textbf{336.73794}	&\textbf{29.22537}	&\textbf{19.6}	&\textbf{2012.26}	&\textbf{0.050}\\
3&OGLEBG 737206	&270.92908	&-28.42642	&0.171	&15.9	&270.92897	&-28.42691	&19.6	&2012.35	&0.756\\
4&LSPM J2205+3730	&331.44064	&37.50056	&0.279	&16.7	&331.44176	&37.50014	&19.6	&2012.35	&0.198\\
\rowcolor{dunkelgrau}
\textbf{5}&\textbf{OGLEBG 490395}	&\textbf{271.39029}	&\textbf{-28.19142}	&\textbf{0.186}	&\textbf{17.6}	&\textbf{271.39011}	&\textbf{-28.19197}	&\textbf{16.2} &\textbf{2012.45}	&\textbf{0.125}\\
6&LSPM J2104+5325	&316.09170	&53.42085	&0.221	&19.1	&316.09250	&53.42135	&19.8	&2012.49	&0.148\\
7&LSPM J0916+0926	&139.05100	&9.44044	&0.212	&13.0	&139.05022	&9.43994	&18.9	&2012.55	&0.386\\
8&LSPM J0905+6733	&136.43961	&67.56268	&0.862	&14.6	&136.43501	&67.56012	&20.7	&2012.83	&0.405\\
9&LSPM J1905+6022	&286.49097	&60.37719	&0.240	&17.9	&286.49023	&60.37802	&18.4	&2012.88	&0.046\\
\rowcolor{dunkelgrau}
\textbf{10}&\textbf{LSPM J1948+3250}	&\textbf{297.10131}	&\textbf{32.84255}	&\textbf{0.215}	&\textbf{14.9}	&\textbf{297.10186}	&\textbf{32.84325}	&\textbf{17.0}	&\textbf{2012.94}	&\textbf{0.129}\\
11&LSPM J0352+3620	&58.02211	&36.34658	&0.156	&19.6	&58.02267	&36.34599	&20.6	&2012.95	&1.504\\
12&LSPM J1441+1731	&220.28276	&17.52493	&0.228	&18.3	&220.28340	&17.52436	&20.8	&2012.96	&0.496\\
\hline\hline
13&LSPM J0427+7609	&66.87036	&76.16094	&0.342	&14.6	&66.87380	&76.15995	&20.1	&2013.16	&0.234\\
14&LSPM J1943+0941	&295.80650	&9.69005	&0.507	&17.2	&295.80546	&9.68842	&16.1	&2013.62	&0.675\\
15&LSPM J0020+0044	&5.20582	&0.74310	&0.207	&17.0	&5.20656	&0.74287	&20.3	&2013.76	&1.268\\
16&PPMX 130002.0	&195.00840	&-28.72489	&0.431	&14.8	&195.00987	&-28.72590	&19.4	&2013.79	&0.070\\
&~~~~~~-284329	&	&	&	&	&	&	&	&	&\\
17&LSPM J2111+3123	&317.86120	&31.38924	&0.235	&18.1	&317.86223	&31.38967	&19.9	&2013.95	&0.481\\
\hline\hline
\rowcolor{dunkelgrau}
\textbf{18}&\textbf{LSPM J0431+5858E}	&\textbf{67.80238}	&\textbf{58.97810}	&\textbf{2.375}	&\textbf{12.1}	&\textbf{67.81246}	&\textbf{58.97045}	&\textbf{19.7}	&\textbf{2014.04}	&\textbf{0.133}\\
\rowcolor{dunkelgrau}
\textbf{19}&\textbf{LSPM J0207+4938}	&\textbf{31.76613}	&\textbf{49.64538}	&\textbf{0.486}	&\textbf{12.1}	&\textbf{31.76761}	&\textbf{49.64367}	&\textbf{18.7}	&\textbf{2014.34}	&\textbf{0.061}\\
\rowcolor{dunkelgrau}
\textbf{20}&\textbf{LSPM J2004+3808}	&\textbf{301.08704}	&\textbf{38.14137}	&\textbf{0.341}	&\textbf{12.7}	&\textbf{301.08606}	&\textbf{38.14012}	&\textbf{12.7}	&\textbf{2014.51}	&\textbf{0.037}\\
21&LSPM J2130+4842	&322.58025	&48.70194	&0.260	&16.0	&322.58137	&48.70279	&19.2	&2014.69	&0.490\\
22&UCAC3  211.75498	&211.75498	&-72.81659	&0.242	&14.7	&211.75163	&-72.81717	&19.6	&2014.72	&0.238\\
&~~~~~~-72.81659	&	&	&	&	&	&	&	&	&\\
23&LSPM J0730+3248	&112.53480	&32.80781	&0.213	&18.3	&112.53393	&32.80738	&17.0	&2014.96	&0.722\\
\hline\hline
24&LSPM J0225+4227	&36.42162	&42.45195	&0.180	&18.2	&36.42300	&42.45174	&17.9	&2015.17	&1.483\\
25&LSPM J1356+2858	&209.14762	&28.98275	&0.188	&14.7	&209.14694	&28.98214	&20.3	&2015.47	&1.331\\
26&LSPM J2035+6453	&308.87289	&64.89700	&0.303	&15.0	&308.87534	&64.89790	&19.2	&2015.49	&1.192\\
27&LSPM J0218+3731	&34.64142	&37.51808	&0.169	&16.3	&34.64229	&37.51820	&20.4	&2015.58	&0.253\\
\rowcolor{dunkelgrau}
\textbf{28}&\textbf{LSPM J2022+2657}	&\textbf{305.61134}	&\textbf{26.95245}	&\textbf{0.331}	&\textbf{16.3}	&\textbf{305.61082}	&\textbf{26.95104}	&\textbf{19.6}	&\textbf{2015.97}	&\textbf{0.012}\\
\hline\hline
29&LSPM J1625+1540	&246.30813	&15.68172	&1.216	&14.2	&246.30872	&15.67635	&20.6	&2016.11	&0.055\\
30&LSPM J1020+2915	&155.07403	&29.25383	&0.289	&19.1	&155.07369	&29.25261	&18.5	&2016.33	&0.514\\
31&LSPM J1330+1909	&202.62941	&19.15944	&1.385	&15.2	&202.62699	&19.15347	&20.3	&2016.75	&0.100\\
32&LSPM J0301+7310	&45.29591	&73.16964	&0.291	&18.8	&45.30031	&73.16911	&19.3	&2016.8	&0.179\\
33&UCAC3  226.36393	&226.36393	&-46.33785	&0.530	&14.7	&226.36077	&-46.33899	&18.7	&2016.9	&0.424\\
&~~~~~~-46.33785	&	&	&	&	&	&	&	&	&\\
\hline\hline
\rowcolor{dunkelgrau}
\textbf{34}&\textbf{LSPM J1209+0042}	&\textbf{182.36808}	&\textbf{0.70392}	&\textbf{0.386}	&\textbf{14.4}	&\textbf{182.36629}	&\textbf{0.70379}	&\textbf{17.9}	&\textbf{2016.95}	&\textbf{0.152}\\
35&LSPM J1502+3531	&225.54918	&35.53155	&0.411	&19.3	&225.54721	&35.53039	&19.8	&2017.28	&0.699\\
36&LSPM J2008+2358	&302.11758	&23.96755	&0.220	&18.3	&302.11870	&23.96806	&19.6	&2017.42	&0.724\\
37&LSPM J0107+3412	&16.94937	&34.20847	&1.462	&12.8	&16.95775	&34.21086	&19.7	&2017.46	&0.485\\
38&UCAC3  253.86047	&253.86047	&-64.54973	&0.203	&15.9	&253.85828	&-64.55027	&18.6	&2017.68	&0.226\\
&~~~~~~-64.54973	&	&	&	&	&	&	&	&	&\\
\hline\hline
39&LSPM J0646+1304	&101.58121	&13.07978	&0.255	&18.1	&101.58178	&13.07863	&18.9	&2018.29	&0.054\\
40&LSPM J0729+3308	&112.45441	&33.14243	&0.191	&19.0	&112.45431	&33.14201	&19.2	&2018.34	&1.522\\
\hline\hline
41&LSPM J0228+6553	&37.01399	&65.89983	&0.192	&19.3	&37.01595	&65.89915	&20.1	&2019.06	&0.352\\
42&LSPM J0633+5257	&98.36513	&52.96467	&0.198	&19.9	&98.36574	&52.96358	&20.6	&2019.51	&0.389\\
43&LSPM J0450+1704	&72.66384	&17.08175	&0.216	&15.1	&72.66481	&17.08098	&16.9	&2019.88	&0.328\\
\hline
\end{tabular}
\captionsetup{format=plain}\singlespacing \vspace{-0.3cm}
\caption{The 43 real (out of 1118 supposed) astrometric microlensing candidates between 2012 and 2019. Given are the lens ID, the lens and source coordinates at equinox J2000.0, the apparent magnitudes of lens and source, the lens proper motion, the time at closest approach, and the minimal angular distance of lens and source. 9 of these candidates are printed in boldface and grey background because they could lead to an observable astrometric microlensing event. Detailed information of these 9 expected astrometric microlensing events are given in Table 2.}
\end{table}

\onecolumn
\begin{figure}[t]
	\centering
	\includegraphics[width=15cm]{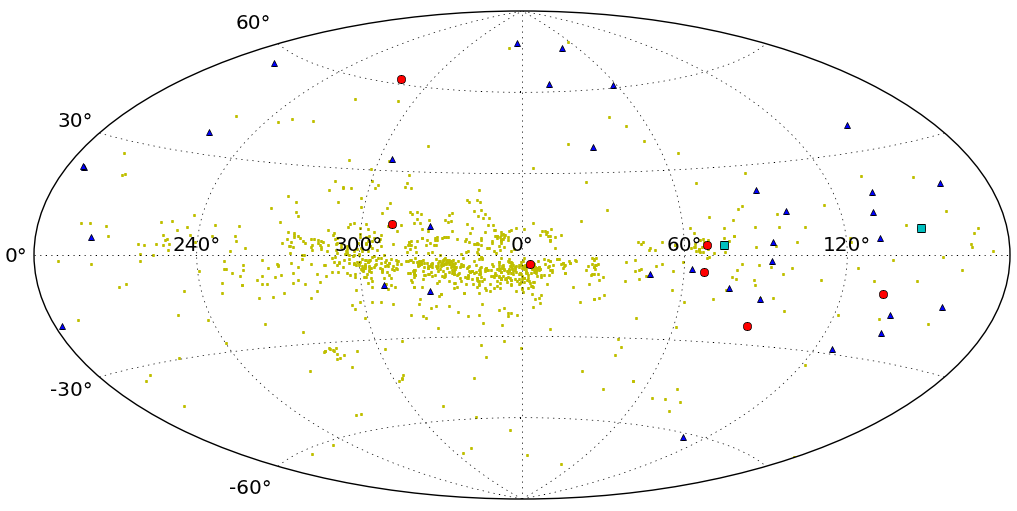}
	\captionsetup{format=plain}\vspace{0.2cm}
	\caption{The distribution of all 1118 supposed microlensing candidates. The 1075 spurious candidates (yellow dots), the 43 real candidates (blue triangles, red circles, and green squares) and the 9 best microlensing events (red circles and green squares) are represented in Galactic coordinates (Aitoff projection). The 2 events that should be observable with Gaia are plotted as green squares. The majority of microlensing candidates and possible events are close to the Galactic plane. }
	\label{fig::Aitoff}
\twocolumn
\end{figure}

\noindent
spectral type and the luminosity class. For the White Dwarf masses we used the mean value of the mass-distribution from DA-White Dwarfs from the Palomar Green Survey \citep{astro-ph/0406657v2}, since the majority of White Dwarfs are in this group. The analysis yields that all lens masses are smaller than the mass of the sun (0.08 to 0.6 solar masses).

For the sources, a distance determination was not possible. Hence we had to use an estimation of the Einstein radius for distant sources, which is
\begin{equation}
	\theta_\text{E}=0.902\text{mas}~\sqrt{\frac{M}{M_\odot}\frac{10\text{kpc}}{D_\text{L}}}.
\end{equation}
In this way the Einstein radii are estimated to between 2 and 30 mas. The minimal impact parameters of our lensing candidates are $\sim$2 to $\sim$600 Einstein radii. Assuming a dark lens, we would get maximal centroid shifts of $\sim$4 to $\sim$6000 µas. But when the lens brightness is equal or larger than the source brightness we have to consider this effect. Unfortunately, this is the case for the most candidates at closest approach. Thus we calculated the maximal centroid shift for a luminous lens with the assumption of large impact parameters (equation \ref{equ::CenShiftMaxLum}). Besides we determined the centroid shift at the projected lens-source distance of 200 mas (Gaia's angular resolution) where the luminous lens effect has not to be considered anymore ($\delta\theta_\text{c,res}$).

The time interval, for which a large enough astrometric signal is measurable with Gaia, is defined as $t_\text{ast}$ (equation 15). For our 43 microlensing candidates, this time interval ranges between 3 days and 2525 days.

After analysis of the source characteristics we could identify only one extended source. This source is probably a galaxy and not a pointlike object thus the corresponding lens is not a suitable candidate (\#16 in Table 1). 

Now we want to identify the best candidates using the above criteria. The study of the maximal centroid shift ($\delta\theta_\text{c}$ at $u_0$ for $u_0\geq\sqrt{2}$), considering the precision compared to the threshold $\theta_\text{min}$, yields 9 astrometric microlensing candidates with closest 
\\\\\\\\

\vspace{8.83cm}
\noindent
approach between 2012 and 2019. Due to positional uncertainties, proper motion uncertainties and uncertainties of the estimated masses and distances, the accuracy of the centroid shift has the same order of magnitude as the centroid shift itself for an individual measurement. When we include the luminous lens effect and the above criteria only 2 very good candidates remain that fulfill all required criteria. This means that under ideal conditions it is possible to measure the microlensing effect for 9 candidates with Gaia, but it appears robust only for 2 of them.

\vspace{0.2cm}
The 9 possible astrometric microlensing candidates, their positions and proper motions, and calculated parameters are shown in Table 2, ordered by time of closest approach. In this table the 2 best microlensing candidates are \#18 and \#20.

\vspace{0.2cm}
The Galactic distribution of all microlensing candidates is shown in Figure \ref{fig::Aitoff}. The yellow dots represent the 1075 spurious candidates, the other symbols (blue triangles, red circles, green squares) the 43 real microlensing candidates, the red circles and green squares the 9 best candidates, and the green squares the 2 probably measurable microlensing events. Due to a higher density of background stars, the majority of microlensing candidates are close to the Galactic plane. Because of crowding this is also the region with the highest rate of spurious high proper motions and hence the part of the sky that contains most erroneous candidates. 

\subsection{Excellent Astrometric Microlensing Candidate LSPM J0431+5858E}

Now we will give some more information on the 2 best lensing candidates. The first candidate is the White Dwarf LSPM J0431+5858E (also known as GJ 169.1 B, LHS 27 or Stein 2051 B) which is a member of a double star system. The projected binary separation is about 8''. It is a very widely separated system and hence the effect of the second lens, which has probably only half the mass of the primary, is negligible. With equation (\ref{equ::deltaTheta2}) it can be seen that the resulting centroid shift of the second compo-
\begin{landscape}
\begin{table}\setstretch{1.15}
\centering\vspace{2.5cm}
\begin{tabular}{|l|l|l|l|l|>{\columncolor{dunkelgrau}}l|l|>{\columncolor{dunkelgrau}}l|l|l|} 
\hline
\#&1&2&5&10&\textbf{18}&19&\textbf{20}&28&34\\\hline
lens ID				&UCAC3  206.42		&LSPM J2226	&OGLEBG		&LSPM J1948	&\textbf{LSPM J0431}	&LSPM J0207	&\textbf{LSPM J2004}	&LSPM J2022	&LSPM J1209\\
				&273-051.01684 		&+2913		&490395		&+3250		& \textbf{+5858E}	&+4938		&\textbf{+3808}		&+2657	&+0042	\\\hline\hline
$t_\text{min}$ (yr)		&2012.11$\pm$0.97	&2012.26$\pm$0.65	&2012.45$\pm$6.39	&2012.94$\pm$0.66	&\textbf{2014.04$\pm$0.08}&2014.34$\pm$0.33	&\textbf{2014.51$\pm$0.60}		&2015.97$\pm$0.56	&2016.95$\pm$0.36\\
$\vartheta_\text{min}$ ('')		&0.070$\pm$0.333	&0.050$\pm$0.194	&0.125$\pm$0.468	&0.129$\pm$0.142	&\textbf{0.133$\pm$0.160}&0.061$\pm$0.130	&\textbf{0.037$\pm$0.201}		&0.012$\pm$0.173	&0.152$\pm$0.137\\
$l$ (deg)			&311.57425		&88.69476		&2.84945		&68.24745		&\textbf{148.10341}&135.29101		&\textbf{74.47165}			&67.22408		&280.33137\\
$b$ (deg)			&10.94226		&-23.87574		&-3.40778		&3.63157		&\textbf{7.31408}&-11.40291		&\textbf{3.59330}			&-5.74691		&61.72031\\
$\mu_\text{L}$ (''/yr)		&0.527			&0.230			&0.186			&0.215			&\textbf{2.375}&0.486			&\textbf{0.341}				&0.331			&0.386\\
$V_\text{L}$ (mag)		&15.3			&13.6			&17.6			&14.9			&\textbf{12.1}&12.1			&\textbf{12.7}				&16.3			&14.4\\
MK				&M4 V			&M4 V			&M4 V			&K7 V			&\textbf{DC WD}&M3 V			&\textbf{M1 V}				&M4 V			&M3 V\\
$M_\text{V}$ (mag)		&11.5			&11.5			&11.5			&8.5			&\textbf{13.3}&10.7			&\textbf{9.5}				&11.5			&10.7\\
$M_\text{L}$ (M$_\odot$) 	&0.3			&0.3			&0.3			&0.55			&\textbf{0.6}&0.35			&\textbf{0.45}				&0.3			&0.35\\
$D_\text{L}$ (pc)		&57.7	&26.1	&169.3	&192.3	&\textbf{5.6}&19.3			&\textbf{42.9}				&92.5			&55.7\\
$\theta_\text{E}$ (mas)		&6.507	&9.678	&3.797	&4.824	&\textbf{29.560}&12.141			&\textbf{9.243}				&5.138			&7.149\\
$u_0$ ($\theta_\text{E}$)	&10.7	&5.2	&32.9	&26.7	&\textbf{4.5}&5.0			&\textbf{4.0}				&2.4			&21.3\\
$\sigma_\text{a}$ (µas)		&540	&1050	&150	&230	&\textbf{1100}&610			&\textbf{30}				&1050			&380\\
$\theta_\text{min}$ (µas)	&1025	&1992	&285	&436	&\textbf{2087}&1157			&\textbf{57}				&1992			&721\\
$\delta\theta_\text{c,max}$ (µas)	&597$\pm$2764	&1733$\pm$5801	&115$\pm$434	&180$\pm$216	&\textbf{5972$\pm$5946}&2239$\pm$4140		&\textbf{2069$\pm$8810}			&1591$\pm$10782		&335$\pm$343\\
$\delta\theta_\text{c,res}$ (µas)	&211$\pm$88	&466$\pm$193	&72$\pm$34	&116$\pm$55	&\textbf{4186$\pm$636}&732$\pm$278		&\textbf{425$\pm$153}			&132$\pm$46		&255$\pm$128\\
$\delta\theta_\text{c,lum}$ (µas)	&30$\pm$147	&7$\pm$29	&92$\pm$346	&22$\pm$29	&\textbf{6$\pm$8}&5$\pm$12		&\textbf{1083$\pm$6332}			&93$\pm$1368		&13$\pm$15\\
$t_\text{E}$ (days)		&5	&15	&7	&8	&\textbf{5}&9			&\textbf{10}				&6			&7\\
$t_\text{ast}$ (days)		&45	&117	&156	&142	&\textbf{101}&150			&\textbf{2525}				&23			&105\\
$\mu_S$ (''/yr)			&0.002	&0.038	&0.029	&0.026	&\textbf{0.011}  &0.007			&\textbf{0.024}				&0.007			&0.010\\
$V_\text{S}$ (mag)		&18.5	&19.6	&16.2	&17.0	&\textbf{19.7}&18.7			&\textbf{12.7}				&19.6			&17.9\\
\hline
\end{tabular}
\captionsetup{format=plain}\singlespacing \vspace{-0.2cm}
\caption{List of the best astrometric microlensing candidates with closest approach between 2012 and 2019 that could be observable within the Gaia mission. The numbers are those of Table 1. The boldface candidates 18 and 20 should be easily measurable with Gaia. The following quantities are shown: time of closest approach ($t_\text{min}$), minimal angular distance ($\vartheta_\text{min}$), Galactic coordinates of the lens (longitude $l$, latitude $b$), proper motions of lens and source ($\mu_\text{L}$; $\mu_\text{S}$), apparent magnitudes ($V_\text{L}$; $V_\text{S}$), lens spectral type (MK), absolute magnitude of the lens ($M_\text{V}$), lens mass ($M_\text{L}$), lens distance ($D_\text{L}$), angular Einstein radius ($\theta_\text{E}$), minimal impact parameter ($u_0$), positional accuracy ($\sigma_\text{a}$), accuracy threshold for five consecutive observations ($\theta_\text{min}$), maximal centroid shift for a dark lens ($\delta\theta_\text{c,max}$), maximal centroid shift with luminous lens effect ($\delta\theta_\text{c,lum}$), centroid shift at $\vartheta=$200 mas ($\delta\theta_\text{c,res}$), Einstein time ($t_\text{E}$) and astrometric event duration ($t_\text{ast}$).}
\end{table}
\end{landscape}
\noindent
nent would be only 2 µas. The lens has a distance to the sun of only 5.6 pc and the highest maximal centroid shift (dark lens) of all 43 candidates ($\approx$6000 µas). The source is very faint (19.7 mag) and the lens is much brighter (12.1 mag) hence we have to consider the luminous lens effect. But when lens and source are resolvable the centroid shift is still $\approx$4200 µas ($\theta_\text{min}\approx2100$ µas). The closest approach of lens and source will be on January 2014 ($\pm$1 month). The time $t_\text{ast}$, in which Gaia could measure the astrometric deviation for this event, is about 100 ($\pm$22) days. The mean end-of-mission number of Gaia observations is $\approx$80 spread over about 5 years \citep{ESAPER,2010IAUS..261..296L,2011EAS....45...97M}, hence Gaia should measure a clear centroid shift signal at a few different epochs. 
%

\subsection{Excellent Astrometric Microlensing Candidate LSPM J2004+3808}

The second promising lensing candidate is the M-Dwarf LSPM J2004+3808 (G 125-56) in 43 pc distance. Because lens and source have nearly the same magnitude (12.7 mag), the centroid shift trajectory is less affected by the luminous lens effect. The maximal centroid shift would be $\approx$2070 µas for a dark lens and is expected to be $\approx$1080 µas with consideration of the lens brightness. The accuracy threshold $\theta_\text{min}$ amounts to only 57 µas due to the bright source star. The closest approach of lens and source will be in July 2014 ($\pm$6 months). The calculated possible observing time with Gaia ($t_\text{ast}$) is 2490 ($\pm$1275) days hence the event will be measurable over the whole mission (about 80 times).

\section{Summary \& Outlook} 
Our goal was to find and characterize high proper motion stars as candidates for astrometric microlensing detectable with Gaia. We selected 43 microlensing candidates which are available in our list ''Candidates of Astrometric Lensing''. We used the LSPM, UCAC3, PPMX, and OGLEBG as lens catalogs and the PPMXL as source catalog. The LSPM was a very good selection for the lenses but it is only available for the northern sky. The proper motion analysis with Aladin and SIMBAD showed that 1075 of 1118 (96\%) supposed candidates have erroneous proper motions, but all LSPM lenses (36) have correctly determined proper motions. It is well known that this is a typical problem of high proper motions in star catalogs.

We have investigated maximal centroid shifts (for dark and luminous lenses), Gaia's accuracy threshold $\theta_\text{min}$, the type of source and the average astrometric event duration to determine which of the 43 remaining actual lensing candidates are detectable with Gaia. Our analysis yields 9 astrometric lensing candidates that could be measurable with Gaia. However, for many of them the observability is marginal due to large centroid shift errors and too short event durations. Two of them have a very strong centroid shift and a long event duration and hence they should be detectable, perhaps also with ground based telescopes astrometrically. These 2 candidates should also lead to a photometrically measurable signal of some millimag. With an Einstein time of 4.5 and 9.9 days, this photometric signal is hardly observable with Gaia, but we can monitor them photometrically from the ground. It is not possible to estimate whether the other 7 possible candidates are observable photometrically due to very short photometric event durations, large impact parameters, and large impact parameter uncertainties , i.e. large uncertainties in the magnitude shift (equation \ref{equ::deltaMag}). 

Gaia's data will be used to produce a catalog with very high precise positions, proper motions and magnitudes of about one billion stars and other astronomical objects with visual magnitudes up to 20. This dataset can then be used for an improvement of the microlensing catalog. On this basis one can predict possible lensing events with much higher accuracy than today. These events could be observed astrometrically with interferometry and photometrically. Hence many single star masses in the solar neighborhood will be determined in the not-so-distant future. The proper motion determinations of Gaia will be very precise so that it would not be necessary to verify all proper motions from the predicted lensing candidates.

\begin{acknowledgements}
It is a pleasure to thank Mark Taylor for providing the Tool for Operations on Catalogues And Tables TOPCAT \citep{2005ASPC..347...29T} which was used to prepare five figures in this paper. Besides we wish to thank the members of the gravitational lensing group at ARI for helpful comments and discussions to our work, Ulrich Bastian and Stefan Jordan for information about the Gaia mission, Siegfried Roeser for information about astronomical catalogs and their problems, and Kailash Sahu for helpful suggestions on the manuscript. We would like to particularly thank the referee, Andy Gould, for very constructive comments which helped to improve the manuscript significantly.

This publication makes use of data products from the Two Micron All Sky Survey, which is a joint project of the University of Massachusetts and the Infrared Processing and Analysis Center/California Institute of Technology, funded by the National Aeronautics and Space Administration and the National Science Foundation.
\end{acknowledgements}

	\begin{figure*}[tbp]
		\centering
		\includegraphics[width=0.8\textwidth]{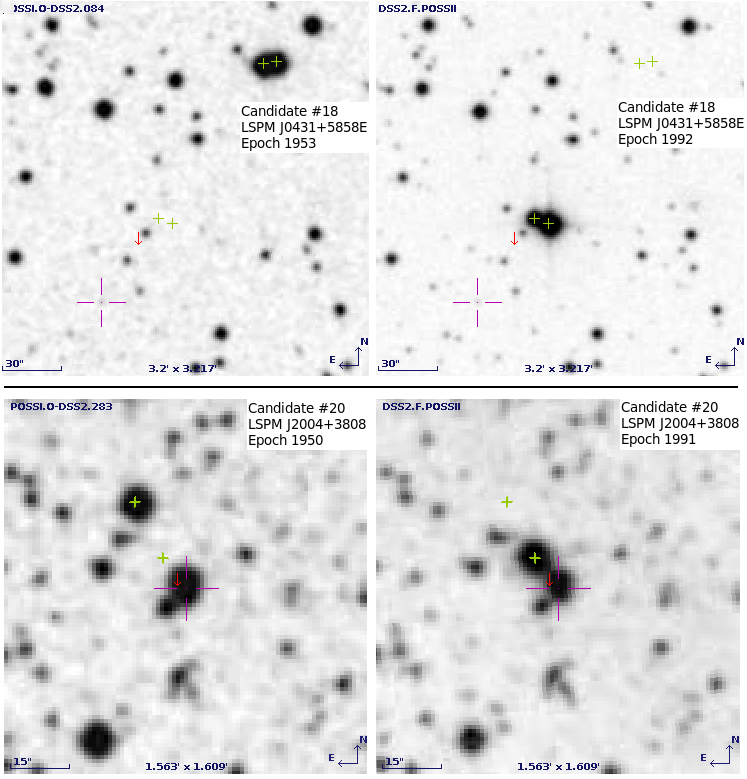}
  		\captionsetup{format=plain}
		\caption{\label{abb::PaperEvents}Images from the POSSI and POSSII survey for the 2 best astrometric microlensing candidates produced with Aladin. The upper two figures show the lens candidate LSPM J0431+5858E (candidate \#18). The lens candidate is a White Dwarf with an estimated mass of 0.6M$_\odot$ in 5.6 pc distance and it is a member of a double star system. The two pairs of green plus symbols indicate the double system in the years 1953 and 1992 (the left star is our lens) and the red arrow presents the calculated lens position at the epoch 2000.0. On the bottom panels one can see the second best lensing candidate LSPM J2004+3808 (candidate \#20) which has a distance of about 43 pc to the sun. The lens positions in the years 1950 and 1991 are also marked with green plus symbols as well as its position at 2000 is marked with a red arrow. In all images the source star is denoted with a purple cross. One can see that the proper motions of the 2 lensing candidates are correctly given in the LSPM-NORTH catalog.}
\end{figure*}

\bibliographystyle{aa} 
\bibliography{Paper}

\end{document}